\documentclass[conference]{IEEEtran}
\IEEEoverridecommandlockouts
\usepackage{cite}
\usepackage{amsmath,amssymb,amsfonts}
\usepackage{mathtools}
\usepackage{algorithmic}
\usepackage{graphicx}
\usepackage{textcomp}
\usepackage{tikz}
\usepackage{subcaption}
\usepackage{soul}
\def\BibTeX{{\rm B\kern-.05em{\sc i\kern-.025em b}\kern-.08em
    T\kern-.1667em\lower.7ex\hbox{E}\kern-.125emX}}

\usepackage[dvipsnames,svgnames,table]{xcolor}
\usepackage[linecolor=DarkGreen,backgroundcolor=DarkGreen!25,bordercolor=DarkGreen,textsize=tiny]{todonotes}

\usepackage{url}
\usepackage{makecell}
\usepackage{caption}

\usepackage{float}

\usepackage{multirow}
\usepackage{lipsum}
\usepackage{wrapfig}

\usepackage{xspace}
\usepackage{pifont}
\usepackage{listings}

\usepackage{booktabs,tabularx,array,amssymb}
\newcolumntype{L}{>{\raggedright\arraybackslash}X}
\newcolumntype{C}[1]{>{\centering\arraybackslash}p{#1}}

\newcommand*\circled[1]{\tikz[baseline=(char.base)]{
            \node[shape=circle,fill,inner sep=1pt,scale=0.8] (char) {\textcolor{white}{#1}};}}

\usepackage{siunitx}

\usepackage{adjustbox}
\usepackage{bm}
\usepackage{soul}

\newcommand{\name}{\emph{TaxBreak}\xspace}

{\color{red}}%
{}

\usepackage{tcolorbox}
\tcbuselibrary{skins,breakable}
\newtcolorbox{codebox}{
  enhanced,
  colback=gray!5,
  colframe=gray!60,
  boxrule=0.4pt,
  arc=2pt,
  outer arc=2pt,
  left=4pt,
  right=4pt,
  top=4pt,
  bottom=4pt,
  fontupper=\ttfamily\scriptsize,
  breakable
}

\usepackage{forest}
\usetikzlibrary{arrows.meta}

\newcommand{\cmark}{\textcolor{green!60!black}{\ding{51}}} % ✓
\newcommand{\xmark}{\textcolor{red!70!black}{\ding{55}}}   % ✗

\begin{document}

\title{\name: Unmasking the Hidden Costs of LLM Inference Through Overhead Decomposition}

\author{\IEEEauthorblockN{
Prabhu Vellaisamy, Shreesh Tripathi, Vignesh Natarajan, Surya Santhan Thenarasu, Shawn Blanton, John P. Shen}
\IEEEauthorblockA{
\textit{NeuroAI Computer Architecture Lab (NCAL)} \\
\textit{Electrical and Computer Engineering Department} \\
Carnegie Mellon University \\
(pvellais@andrew.cmu.edu, jpshen@cmu.edu)
}}

\maketitle

\begin{abstract}
Large Language Model (LLM) inference is widely used in interactive assistants and agentic systems. In latency-sensitive deployments, inference time can become dominated by host-side overheads. Existing approaches typically expose this cost only as an aggregate residual or a launch/queue metric, which is often insufficient to identify which execution layer should be optimized. This work presents \name, a trace-driven methodology for decomposing host-visible orchestration overhead into three components: framework translation time, CUDA library translation time, and kernel launch-path time. We validate \name on NVIDIA H100 and H200 systems and use it to derive our proposed \textit{Host–Device Balance Index (HDBI)}, a boundedness summary index that relates device-active execution to host-visible orchestration. Across representative dense and mixture-of-experts workloads in both prefill and decode, we show that aggregate latency, GPU inactivity, or boundedness ratios alone can obscure the dominant optimization target. \name instead distinguishes cases where optimization should reduce software-stack overhead from cases where the primary win comes from reducing device-side work. We further show that MoE models dispatch 8-11$\times$ more kernels per output token than dense models, and that for such host-bound workloads, CPU single-thread performance is a first-order parameter: a faster host CPU reduces orchestration overhead by 10-29\% and improves end-to-end latency by up to 14\%, even when paired with a slower-clocked GPU. These results position \name as a diagnostic tool for assessing whether optimization effort should target the software stack or the device-side workload execution.

\end{abstract}


\section{Introduction}

Large language models (LLMs), spanning dense (e.g., GPT-4~\cite{achiam2023gpt}, LLaMA~\cite{grattafiori2024llama, touvron2023llama}) and mixture-of-experts (MoE) (e.g., DeepSeek-V3, R1~\cite{liu2024deepseek_v3, guo2025deepseek}) architectures, now underpin interactive AI services operating under strict latency constraints~\cite{narayanan2021efficient}. In user-facing systems such as chatbots and coding assistants, end-to-end delay directly impacts user experience~\cite{chen2021evaluating, weber2013improving}. Agentic workflows amplify this sensitivity: interleaving reasoning steps with tool execution introduces sequential dependencies, causing overhead to accumulate across iterations~\cite{yao2022react, schick2023toolformer, shinn2023reflexion}. Recent work indicates that, in such settings, host-side orchestration increasingly lies on the critical path, especially when control flow and tool invocation occur outside the GPU execution stream~\cite{raj2025cpu}. Critically, in eager mode, PyTorch dispatches the entire path serially on a single CPU thread, so the relevant CPU performance characteristic is single-core throughput, not core count.

\begin{figure}[!h]
  \centering
  \includegraphics[width=0.51\textwidth, height=4.2cm]{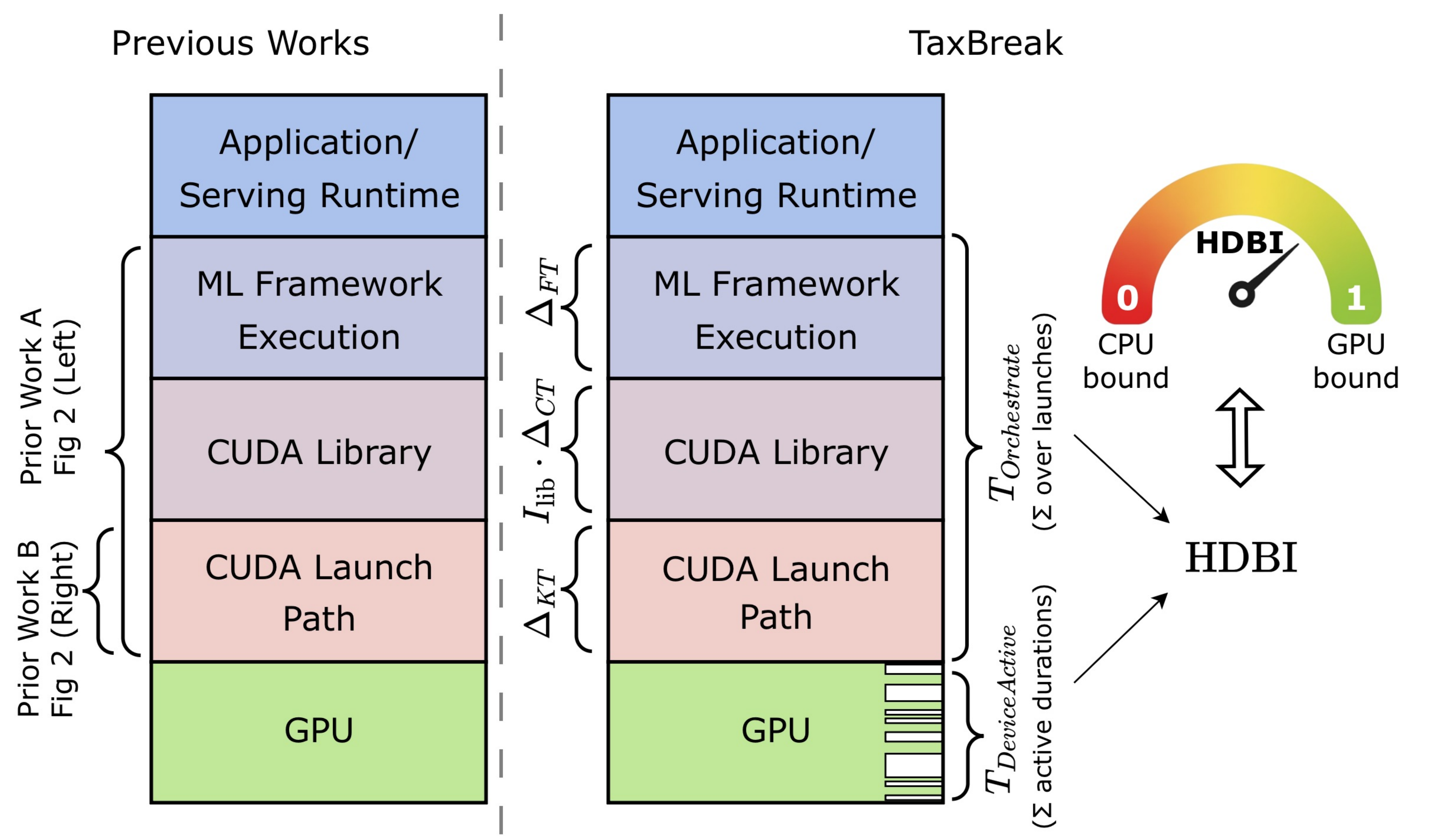}
  \caption{\textbf{\name Methodology.} \name decomposes overall host-side orchestration overhead into three components: (i) framework execution, (ii) CUDA library front-end execution, and (iii) kernel launch invocation. We also introduce a new \emph{Host-Device Balance Index (HDBI)} to characterize the relative boundedness between the host (CPU) and the device (GPU). Prior work~A refers to aggregate framework tax~\cite{fernandez2023framework}; B refers to kernel launch/queue tax (TKLQT)~\cite{vellaisamy2025characterizing}.}
    \vspace{-10pt}
  \label{fig:overview}
\end{figure}

Today, LLM inference is run via deeply layered inference stacks. High-level runtimes (vLLM~\cite{kwon2023efficient}, Orca~\cite{yu2022orca}) manage batching, scheduling, and KV-cache state before dispatching execution through ML frameworks or compilers (PyTorch/TensorFlow~\cite{paszke2019pytorch, pang2020tensorflow}, TVM~\cite{chen2018tvm}). These, in turn, may or may not invoke vendor libraries (cuBLAS/cuBLASLt/cuDNN) and submit kernels. Although this modularity enables portability and abstraction, it introduces host-side orchestration overhead across multiple layers. Prior work typically exposes this cost either as an aggregated CPU-side residual (the \emph{framework tax}~\cite{fernandez2023framework}), or by isolating kernel launch/queue behavior (the \emph{total kernel launch and queue time (TKLQT)}~\cite{vellaisamy2025characterizing}). These metrics are useful for identifying whether a workload appears host- or device-limited, but they do not pinpoint which layer of the stack is the dominant impediment to be optimized.

The attribution challenge intensifies for modern workloads. Inference-centric studies show that performance reflects both hardware capability and host software orchestration across heterogeneous systems~\cite{vellaisamy2025characterizing, li2024large}. GPU-focused analyses further demonstrate strong phase dependence (prefill vs. decode) and scaling effects invisible to coarse-grain profiling~\cite{wang2025systematic, davies2025liminal}. MoE architectures add complexity: although they achieve high effective capacity through sparse activation (e.g., DeepSeek-V3 activates 37B of 671B parameters per token~\cite{liu2024deepseek_v3}), they incur dynamic routing overhead and fragment execution into many small, dependent kernels~\cite{huang2024harder, aimuyo2025flashdmoe}. Even in non-distributed settings, elevated GPU idle time may stem from host-induced delays, workload-induced fragmentation, or memory/data-movement effects, yet existing profilers lack the cross-stack granularity needed to distinguish these causes.

There is a need for a diagnostic framework that attributes inference inefficiencies across host-side abstraction layers and execution phases. To address this, we introduce \name, a trace-driven methodology that decomposes host-side orchestration into three components: (i) \emph{framework execution}, (ii) \emph{CUDA library front-end execution}, and (iii) \emph{kernel launch} targeting NVIDIA systems. We further propose the \emph{Host–Device Balance Index (HDBI)} to quantify relative CPU- vs. GPU-boundedness (Fig.~\ref{fig:overview}). This decomposition turns a coarse host vs. device diagnosis into a mechanism-level attribution, guiding whether optimization should target runtimes, compilers, CUDA/driver pathways, or device-side workload.

\textbf{Key Contributions.}
This work introduces \name, a trace-driven methodology for decomposing host-side LLM inference overhead.
Our contributions include: \circled{1} A three-stage \name decomposition of framework translation, CUDA library translation, and kernel launch overhead; \circled{2} Validation on NVIDIA H100/H200, including per-kernel launch-floor characterization; \circled{3} Empirical evidence across dense and MoE inference in prefill and decode that aggregate metrics by themselves can obscure the dominant optimization target; \circled{4} Host-Device Balance Index (HDBI), a diagnostic summary to complement \name attribution to show if a workload is host- or device-bound; and \circled{5} A cross-platform result showing that faster CPU single-thread performance reduces host-bound end-to-end latency by 11-14\%.

\section{Background and Motivation}

\subsection{Inference Serving Dynamics}
LLM inference serving is divided into two phases: \emph{prefill} and \emph{decode}. During prefill, the server processes the full input prompt to produce the first output token and materializes attention KV cache. Hence, prefill exposes substantial parallelism across prompt tokens and can achieve high GPU utilization when requests are batched. By contrast, the decode phase autoregressively generates tokens one at a time per active request. Decode therefore affords less intra-request parallelism and relies on batching many concurrent requests to improve utilization. Decode is also increasingly shaped by memory behavior because attention repeatedly reads and updates a growing KV cache~\cite{dao2022flashattention}. Production systems thus track time-to-first-token (TTFT) and time-per-output-token (TPOT) as primary latency key performance indicators (KPIs) for interactive responsiveness and steady-state streaming~\cite{agrawal2024taming, agrawal2024vidur}.

The growing memory footprint of KV cache complicates allocation, fragmentation, and reuse, thereby constraining feasible batch sizes and latency. vLLM~\cite{kwon2023efficient} reduces fragmentation by storing the KV state in non-contiguous paged memory, enabling flexible sharing and lower allocation costs. Scheduling approaches such as iteration-level scheduling and selective batching by Orca~\cite{yu2022orca} mitigate inefficiencies from early-finished or late-joining requests. Further, Sarathi-Serve~\cite{agrawal2024taming} targets the throughput–latency tradeoff by leveraging chunked prefills and stall-free scheduling. These serving techniques are typically implemented atop general ML frameworks (PyTorch/TensorFlow/JAX \cite{pang2020tensorflow, paszke2019pytorch, bradbury2021jax}) and vendor runtimes, where framework execution and library/launch paths meaningfully shape TTFT and TPOT. However, the per-layer attribution of those effects remains underexplored and is the focus of this work.

\subsection{Dense and Mixture-of-Experts (MoE) LLMs}

Dense LLMs run a fixed per-token compute graph (all layers active), producing a predictable, largely compute-bound prefill and an iterative decode with fixed per-token overheads and a growing KV cache. In contrast, MoE inference activates a subset of experts per token, adding routing/gating, synchronization, and small, dependency-chained kernels that increase both host-side orchestration (routing dispatch, fragmented kernel submission) and device-side sensitivity (fine-grained dependencies, memory stalls). Coarse metrics (GPU utilization, aggregate launch time) alone cannot distinguish these causes, as observed GPU idle time in MoEs may result from host-induced delays, device dependency chains, or memory/data-movement stalls. Previous studies document routing overhead and imbalance across MoE design points~\cite{chitty2025moe} and show that loss-optimal expert scaling can misalign with serving efficiency~\cite{yun2024toward}, motivating MoE-aware characterization, but not providing the cross-stack attribution that \name provides.

\subsection{Execution Stack Anatomy}

End-to-end LLM latency arises from a layered execution stack spanning application serving logic, framework runtimes, compiler/lowering pipelines, vendor libraries, and submission path. Requests enter a serving runtime (vLLM, Orca) that handles batching, scheduling, and admission control, then invoke model execution through ML frameworks (e.g., PyTorch). In the PyTorch path, the ATen dispatcher resolves tensor metadata and selects backends to target before lowering to vendor libraries (cuBLAS/cuDNN) or directly submitting elementwise and custom kernels to the CUDA launch/submit path.

The PyTorch 2~\cite{ansel2024pytorch} compilation pipeline captures Python execution into FX graphs (TorchDynamo), uses ahead-of-time (AOT) Autograd and optimization passes to fuse and lower operators, and emits backend kernels via TorchInductor. These compilation steps reduce Python overhead but may fall back to eager mode for dynamic workloads. CUDA Graph–based execution can further reduce steady‑state launch overhead, but it incurs a one‑time capture and instantiation cost and typically requires static tensor shapes, memory layouts, and control flow~\cite{vellaisamy2025characterizing}. Nevertheless, MoE routing and expert execution increase both kernel count and heterogeneity, incurring host overhead during decode. Crucially, whether execution is eager or compiled, each inference step still traverses the CUDA runtime/driver submission path: hosts issue runtime/driver calls, work is queued on streams, and the GPU executes kernels. Eager execution issues kernels serially on the host dispatch thread for each request. Although individual launches are asynchronous, the dispatch path remains single-threaded. Consequently, host-side overhead scales linearly with kernel count $N$, and reducing it requires either lowering the per-kernel dispatch cost (faster single-thread execution or compilation) or reducing $N$ directly (kernel fusion, CUDA Graphs), rather than adding CPU cores. Short per-token iterations magnify any fixed per-iteration host overhead. Hence, \name focuses on execution below the framework boundary (PyTorch/ATen via library mapping to the CUDA launch path) and measures the host-side orchestration time. Although compilation is increasingly adopted, eager mode persists for workloads with dynamic shapes or control flow, and \name targets this path.

\begin{figure}[!t]
  \centering
  \includegraphics[width=0.49\textwidth, height=3.6cm]{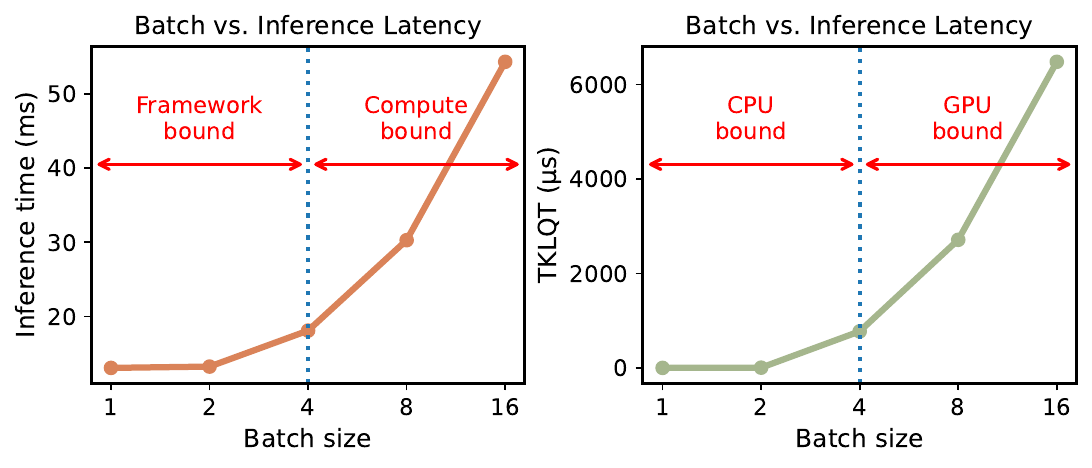}
\caption{\textbf{Previous characterizations of GPT-2 inference across batch sizes.} Left: end-to-end latency (ms) shows a transition from framework-bound to compute-bound at small batch sizes~\cite{fernandez2023framework}. Right: TKLQT in $\mu$s highlights similar CPU-bound to GPU-bound transition as kernel queuing increases with batch size and utilization~\cite{vellaisamy2025characterizing}.}
    \vspace{-10pt}
 % This approach is based on the system launch path.}
  \label{fig:framework_vs_kernel}
\end{figure}

 % We note that \name decomposition relies on the well-defined operator boundaries exposed by PyTorch's eager-mode ATen dispatch. Alternative frameworks or fused-kernel compilers may shift these boundaries.

\subsection{Previous Works on AI Inference Taxes}

Holistic \emph{AI tax} studies show that pipeline and infrastructure costs can dominate real deployments~\cite{richins2021ai, buch2021ai, zhang2025hidden, zhang2024machine}. In server-class LLM inference, the gap between theoretical compute gains and realized latency improvements is referred to as \textit{framework tax}, motivating the framework-bound vs. compute-bound classification~\cite{fernandez2023framework}. CPU-GPU coupling work exposes the kernel launch/queue tax, making kernel submission overhead explicit~\cite{vellaisamy2025characterizing} (Fig.~\ref{fig:framework_vs_kernel}). The kernel launch tax is also relevant to distributed settings where offload and queueing behaviors amplify latency penalties~\cite{trifan2025eliminating}.

\textbf{Limitations of previous works.} \circled{1} Existing studies either present host-side overhead as an aggregate residual~\cite{fernandez2023framework} or localize it mainly to the H2D (host to device) launch path~(TKLQT)~\cite{vellaisamy2025characterizing}. However, neither decomposes pre-launch contributors across framework, library, and driver layers. \circled{2} Prior characterizations are often prefill-centric or target legacy workloads/hardware (e.g., CNNs, encoder models, FP32)~\cite{fernandez2023framework}; device-side analyses focus on microarchitectural behavior but do not attribute host-side stack costs~\cite{wang2025systematic}.

Table \ref{tab:tools} summarizes these comparisons. These gaps motivate us to (i) develop a layer-resolved decomposition separating framework translation, library mapping, and launch-path floor costs; (ii) cover both prefill and decode; and (iii) target dense and MoE workloads on NVIDIA H100/H200.

\begin{table*}[t]
\caption{Comparisons with Previous Works}
\label{tab:tools}
\centering
\footnotesize
\begin{tabularx}{\textwidth}{@{} C{2.2cm} C{2.2cm} C{0.80cm} C{1.2cm} C{4cm} C{1.0cm} C{1.0cm} C{3cm} @{}}
\toprule
\textbf{Work} &
\textbf{Tax Granularity} &
\textbf{CPU-GPU} &
\textbf{Cross-Layer Overhead} &
\textbf{Workloads} &
\textbf{Prefill + Decode} &
\textbf{Hopper-Era HW} &
\textbf{Evaluation Platform(s)} \\
\midrule

AI Tax~\cite{richins2021ai} &
Pipeline-level &
\xmark & \xmark &
Google FaceNet, R-CNN &
\xmark & \xmark &
Edge data center deployment \\

Framework Tax~\cite{fernandez2023framework} &
Coarse residual &
\xmark & \xmark &
BERT-Base; ResNet-50; DistilBERT; GPT-2; WavLM.&
\xmark & \xmark &
NVIDIA Pascal, Turing, Ampere. \\

TKLQT / Kernel-Tax View~\cite{vellaisamy2025characterizing} &
Launch-path only &
\cmark & \xmark &
Bert-Base-Uncased; XLM-Roberta-Base; GPT-2; Llama-3.2-1B. &
\xmark & \cmark &
PCIe A100/H100 and tightly-coupled GH200 (prefill-centric). \\

GPU Inference Characterization~\cite{wang2025systematic} &
Device-centric &
\xmark & \xmark &
Llama-3-8B; Qwen2.5-7B/-32B; Qwen3-30B-A3B. &
\cmark & \xmark &
$4\times$ NVIDIA A100 80GB (SXM) + Jetson AGX Orin. \\

% MoE-Inference-Bench~\cite{chittyvenkata2025moeinferencebench} &
% Model/ops-level &
% \xmark & \xmark &
% MoE LLMs + VLMs (Mixtral/DeepSeek/OLMoE/Qwen families; MoE optimizations). &
% \cmark & \cmark &
% NVIDIA H100 SXM5 80GB. \\

% Inference-Optimal MoE~\cite{yun2024inferenceoptimalmoe} &
% Model-design level &
% \xmark & \xmark &
% MoE LLM scaling/design (inference efficiency as objective; expert-count tradeoffs). &
% \xmark & \xmark &
% Platform not central / not the main contribution. \\

\midrule
\textbf{This Work (TaxBreak)} &
\textbf{Host-stack attribution: $\Delta{FT}$, $\Delta{CT}$, $\Delta{KT}$ } &
\cmark & \cmark &
\textbf{Llama-3.2-1B,-3B, OLMoE-1B/7B; Qwen1.5-MoE-A2.7B} &
\cmark & \cmark &
\textbf{H100 (DGX H100); H200 NVL (single-GPU).} \\
\bottomrule

\end{tabularx}
  \vspace{-8pt}
\end{table*}

%%----------------------------------------------------------------------

\section{\name Methodology}

Before any GPU kernel executes, a PyTorch operation traverses Python dispatch, ATen operator resolution, an optional vendor library front-end, and the CUDA launch API. \name decomposes this host-side latency into three mutually exclusive, collectively exhaustive components per kernel invocation:
\vspace{-10pt}
\begin{multline}
    \label{eq:T_Host}
    T_{Host} = \underbrace{T_{Py} + T_{dispatch\_base}}_{\Delta{FT}} \;+\; \\ \mathbb{I}_{lib} \cdot \underbrace{\max(0,\; T_{dispatch} - T_{dispatch\_base})}_{\Delta{CT}} \;+\; \underbrace{T_{sys}^{floor}}_{\Delta{KT}}
\end{multline}

\noindent where: $\Delta{FT}$ (framework translation) is the sum of Python dispatch overhead $T_{Py} = t_{aten\_op} - t_{torch\_op}$ and the irreducible ATen dispatch cost $T_{dispatch\_base}$; $\Delta{CT}$ (CUDA-Library Translation) is the vendor library front-end excess above that baseline, gated to zero for non-library kernels by $\mathbb{I}_{lib} \in \{0,1\}$; and $\Delta{KT} = T_{sys}^{floor}$ is the hardware floor from \texttt{cudaLaunchKernel} call to GPU kernel start, measured via null-kernel profiling. The raw launch cost is reported separately as $T_{launch}^{raw}=T_{sys}^{floor}+\Delta{KT_{fw}}$ for diagnostic insight, but not used in Eq.~\ref{eq:T_Orchestrate}, as $\Delta{KT_{fw}}$ reflects framework enqueue overhead already captured by $\Delta{FT}$ and $\Delta{CT}$.

Summing over all $N$ kernel invocations in a model run gives the \emph{host-side orchestration overhead}:
\begin{equation}
    \label{eq:T_Orchestrate}
    T_{Orchestration} = \sum_{i=1}^{N}\bigl(\Delta{FT}^{(i)} + \mathbb{I}_{lib}^{(i)} \cdot \Delta{CT}^{(i)} + \Delta{KT}^{(i)}\bigr)
\end{equation}

 Together with device active time aggregated over kernel executions, $T_{DeviceActive} = \sum_{k=1}^{N} t_k$, where $t_k$ denotes the execution time of kernel $k$, these quantities define the \emph{Host-Device Balance Index}:
\begin{equation}
    \label{eq:HDBI}
    HDBI = \frac{T_{DeviceActive}}{T_{DeviceActive} + T_{Orchestration}} \in (0,1) 
\end{equation}
\noindent $HDBI \to 0$ indicates a totally host-bound regime; $HDBI \to 1$ indicates fully device-bound execution. HDBI is a work balance ratio relating device-active time to host orchestration, and not a GPU utilization metric (which measures device-active time against wall-clock latency). Previous work~\cite{fernandez2023framework} defines $T_{Host}$ as inference latency minus GPU active time, but cannot decompose it, while prior kernel-tax analysis~\cite{vellaisamy2025characterizing} reports $\sum\Delta{KT}$ alone. Eq.~\ref{eq:T_Host} provides the full per-kernel decomposition.

\textbf{Diagnostic interpretation using HDBI.} When HDBI signals a host-bound workload ($HDBI \to 0$), the $T_{Orchestration}$ decomposition identifies which layer of the execution stack dominates and thereby suggests an optimization strategy: 

\begin{itemize}
    \item If $\sum\Delta{FT} + \sum\Delta{CT}$ dominates, the bottleneck lies in the software stack (Python dispatch and library front-end overhead), optimizations should target runtime compilation (e.g., \texttt{torch.compile}) or library dispatch paths.
    \item If $N \cdot T_{sys}^{floor}$ dominates, the cost scales with kernel count, and kernel fusion will yield the largest reduction.
    \item If the per-kernel framework launch overhead $\Delta{KT_{fw}} = \max(0,\, T_{launch} - T_{sys}^{floor})$ is significant, the driver/runtime path is the bottleneck, with CUDA~Graphs or persistent kernels helping amortize cost.
\end{itemize}

\subsection{Kernel-Family Taxonomy}

\begin{figure}[t]
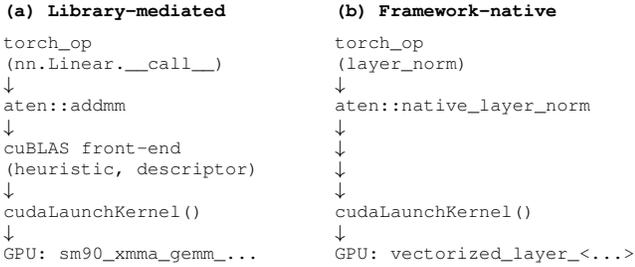

\centering
\scriptsize\ttfamily

\begin{tabular}{p{0.45\linewidth} p{0.45\linewidth}}

\textbf{(a) Library-mediated} &
\textbf{(b) Framework-native}  \\[4pt]

torch\_op &
torch\_op \\

(nn.Linear.\_\_call\_\_) &
(layer\_norm) \\

$\downarrow$ & $\downarrow$ \\

aten::addmm &
aten::native\_layer\_norm \\

$\downarrow$ & $\downarrow$ \\

cuBLAS front-end &
$\downarrow$ \\

(heuristic, descriptor) &
$\downarrow$ \\

$\downarrow$ & $\downarrow$ \\

cudaLaunchKernel() &
cudaLaunchKernel() \\

$\downarrow$ & $\downarrow$ \\

GPU: sm90\_xmma\_gemm\_... &
GPU: vectorized\_layer\_<...> \\

\end{tabular}

\caption{\textbf{Dispatch chains for library-mediated and framework-native kernels.} For library mediated, cuBLAS front-end contributes $\Delta{CT}$ for $\mathbb{I}_{lib}=1$.}
\label{fig:dispatch_tree}

\end{figure}

The $\mathbb{I}_{lib}$ indicator gates the $\Delta{CT}$ term in Eq.~\ref{eq:T_Orchestrate}, ensuring that vendor library overhead is only charged to library-mediated kernels, and determines how a kernel traverses the software stack (illustrated in Fig.~\ref{fig:dispatch_tree}):

\begin{itemize}
\item \textbf{Library-mediated} 
($\mathbb{I}_{lib} = 1$): routed through a vendor library (cuBLAS, cuDNN). Incurs heuristic selection, descriptor setup, and packing before the CUDA launch API, contributing $\Delta{CT} > 0$.

\item \textbf{Framework-native}
($\mathbb{I}_{lib}\!=\!0$): generated directly by ATen/Inductor (e.g., \texttt{aten::mul}, \texttt{aten::native\_layer\_norm}) or data-movement ops (\texttt{cudaMemcpyAsync}, \texttt{cudaMemset}). No vendor library front-end; $\mathbb{I}_{lib}$ gates $\Delta{CT}$ to zero.
\end{itemize}

% \item \textbf{Maintenance and data-movement ($\mathbb{I}_{lib} = 0$)}: Operations such as \texttt{cudaMemcpyAsync}, \texttt{cudaMemset}, and reshape/view kernels.
% \end{itemize}

\subsection{Two-Phase Measurement Pipeline}

\name measures the components of Eq.~\ref{eq:T_Host} in two phases. Phase~1 captures Python-visible framework overhead from a full-model trace and builds a kernel database. Phase~2 first measures the dynamic system floor with a null-kernel \texttt{nsys} run, then replays each unique kernel in isolation to measure dispatch translation, launch-floor cost, and their residuals without queue interference.

\noindent \textbf{Phase 1: Full-model trace.} We execute $R$ profiled iterations after $W$ warm-up runs under PyTorch~Profiler (values in Section~\ref{sec:experiment}), which captures timestamped Python/torch operators, ATen operators, CUDA runtime calls, and GPU kernels linked by correlation IDs. From the last profiled iteration, we extract kernel launch sequences and compute framework translation per invocation:

\begin{equation}
  \Delta{FT}^{(i)} = T_{Py}^{(i)} + T_{dispatch\_base}
\end{equation}

 where $T_{Py}^{(i)} = t_{aten\_op}^{(i)} - t_{torch\_op}^{(i)}$ is the Python-side dispatch overhead before execution reaches the ATen C++ layer, and $T_{dispatch\_base}$ is the irreducible framework dispatch cost measured in Phase~2 from framework-native kernels. We also build a \textit{kernel database} containing each unique kernel's cleaned name, grid/block configuration, ATen metadata, invocation frequency, and $\mathbb{I}_{lib}$ classification.

 \textbf{Phase 2: Isolation replay.} We first measure $T_{sys}^{floor}$ with a null-kernel \texttt{nsys} run. Then, for each kernel-database entry, we replay the original ATen operation in isolation under \texttt{nsys}. Kernels sharing identical ATen metadata (operator, shapes, dtypes, scalar arguments), target kernel name, and launch configuration are deduplicated via a global cache, partitioned so that only uncached entries are profiled, saving significant runtime. Each replay is NVTX-scoped and serialized with \texttt{torch.cuda.synchronize()} to prevent GPU queue overlap (Fig.~\ref{fig:nvtx-timing}).

\vspace{-4pt}

\begin{figure}[!h]
    \centering
    \begin{lstlisting}[language=Python, basicstyle=\small\ttfamily]
nvtx.range_push("aten_dispatch")  # t_nvtx (1)
execute_operation(op_name, inputs) # t_api (2)
nvtx.range_pop()                # t_kernel (3)
torch.cuda.synchronize()     # kernel complete
    \end{lstlisting}
    \caption{\textbf{Annotated NVTX ranges around an operator dispatch and kernel execution.}}
    \label{fig:nvtx-timing}
\end{figure}

After $W$ warm-up iterations, \texttt{nsys} records three timestamps per invocation $j$ (Fig.~\ref{fig:nvtx-timing}): $t_{nvtx}$ at line~(1), $t_{api}$ (\texttt{cudaLaunchKernel}) at line~(2), and $t_{kernel}$ (GPU execution start) at line~(3). For each matched replayed kernel, we report the mean of the per-invocation $T_{dispatch}^{(j)}$ and $T_{launch}^{(j)}$ values over $R$ replay runs. These timestamps map directly to:
\vspace{-1pt}
\begin{equation}
\begin{aligned}
    T_{\text{dispatch}}^{(j)} 
        &= \underbrace{t_{\text{api}}^{(j)} - t_{\text{nvtx}}^{(j)}}_{\text{lines (1)--(2): host dispatch}}
\end{aligned}
\end{equation}
\vspace{-4pt}
\begin{equation}
\begin{aligned}
    T_{\text{launch}}^{(j)} 
        &= \underbrace{t_{\text{kernel}}^{(j)} - t_{\text{api}}^{(j)}}_{\text{lines (2)--(3): launch gap}}
\end{aligned}
\end{equation}

where $t_{nvtx}$, $t_{api}$, and $t_{kernel}$ come from \texttt{NVTX\_EVENTS}, \texttt{CUPTI\_ACTIVITY\_KIND\_RUNTIME}, and \texttt{CUPTI\_ACTIVITY\_KIND\_KERNEL}, respectively, linked by correlation ID.

\subsubsection{Dispatch baseline} The NVTX range encompasses both ATen dispatch and any vendor-library front-end, so $T_{dispatch}$ conflates them for $\mathbb{I}_{lib}{=}1$ kernels. We isolate $\Delta{CT}$ by subtracting the \emph{dispatch baseline}, defined as the median $T_{dispatch}$ of framework-native kernels:
\begin{equation}
    T_{dispatch\_base} = \text{median}\bigl(\{T_{dispatch}^{(k)} : \mathbb{I}_{lib}^{(k)} = 0\}\bigr)
\end{equation}

\noindent We then compute

\vspace{-10pt}
\begin{equation}
    \Delta{CT} = \max(0, T_{dispatch} - T_{dispatch\_base})
\end{equation}

Additionally, $T_{sys}^{floor}$ is the mean $T_{launch}^{raw}$ of an empty C++ \texttt{\_\_global\_\_} null kernel profiled under the same protocol.

\subsubsection{Kernel matching}
Replay may dispatch a variant that differs from the original trace due to autotuning. After narrowing replay candidates to the target neighborhood, we resolve the final kernel via the following name-based fallback hierarchy, where $\bar{n}$ denotes the cleaned (canonical) kernel name:
\begin{equation}
\label{eq:match}
\text{match}(k)=
\begin{cases}
\text{exact}, &
\bar{n}_{\mathrm{replay}}=\bar{n}_{\mathrm{trace}} \\

\text{substring}, &
\begin{aligned}
\bar{n}_{\mathrm{replay}}\subseteq\bar{n}_{\mathrm{trace}}\\
\vee\;
\bar{n}_{\mathrm{trace}}\subseteq\bar{n}_{\mathrm{replay}}
\end{aligned} \\

\text{most-frequent}, &
\text{otherwise}
\end{cases}
\end{equation}

\vspace{-0.22cm}

\section{Experimentation Infrastructure}
\label{sec:experiment}

\subsection{Hardware Setup}

We evaluate on the following NVIDIA GPU platforms:

\begin{itemize}
\item \textbf{H100 platform:} Intel Xeon 8480C (PCIe Gen5, 56 cores @ 2.0/3.8 GHz) paired with an NVIDIA H100 (80\,GB) GPU in a DGX H100 system.
\item \textbf{H200 platform:} Intel Xeon Gold 6538Y+ (32 cores) paired with an NVIDIA H200 NVL (141\,GB) GPU.
\end{itemize}

All experiments use a \textbf{single-GPU Slurm allocation}. To provide a fair comparison across platforms, we allocate \textbf{6 CPU cores} and \textbf{32\,GB} of host memory per GPU on both systems. The 6-core allocation exceeds the single-threaded dispatch path's needs and provides OS scheduling headroom.

% \begin{table}[!h]
%   \centering
%   \caption{Per-GPU host resource allocation.}
%   \label{tab:exp_setup}
%   \begin{tabular}{lcc}
%     \toprule
%     \textbf{Platform} & \textbf{CPU cores per GPU} & \textbf{Host memory per GPU (GB)} \\
%     \midrule
%     H100  & 6 & 32 \\
%     H200  & 6 & 32 \\
%     \bottomrule
%   \end{tabular}
% \vspace{-8pt}
% \end{table}

\textbf{Measurement}. To eliminate cold-start and transient effects, each configuration runs ($W$=50) warm-up iterations followed by ($R$=150) measured runs, for both Phase~1 and Phase~2. Reported latencies are averaged across runs, and the 95\% confidence interval of $T_{Orchestration}$ remains below 0.34~ms across all configurations, indicating stable measurements.

\subsection{Software Setup}

The software setup uses Python 3.13, PyTorch 2.10, CUDA toolkit 12.6 for runtime libraries and NVIDIA Nsight Systems 12.8.1 (\texttt{nsys}) for kernel-level tracing. These traces provide the boundaries needed for the \name decomposition (framework dispatch, CUDA-library front-end processing, and launch-path intervals). Unless otherwise specified, experiments run in eager mode.

\subsection{Workloads}

We profile dense and MoE LLMs in BFloat16, including Llama-3.2-1B/-3B~\cite{grattafiori2024llama} \textbf{(dense)}, GPT-2 (124M), and OLMoE-1B/7B, Qwen1.5-MoE-A2.7B \textbf{(MoEs)}. GPT-2 is used for direct comparison with prior TKLQT characterization~\cite{vellaisamy2025characterizing}.

\section{Experimental Results and Analysis}
\label{sec:results}

\begin{figure*}[t!]
\centering

% ================= Column Headers =================
\makebox[0.249\textwidth][c]{\small\textbf{H100 - Prefill ($m=1$)}}%
\makebox[0.249\textwidth][c]{\small\textbf{H100 - Decode ($m=10$)}}%
\makebox[0.249\textwidth][c]{\small\textbf{H200 - Prefill ($m=1$)}}%
\makebox[0.249\textwidth][c]{\small\textbf{H200 - Decode ($m=10$)}}

%\vspace{4pt}

% ================= Model Label =================
{\footnotesize\textbf{Llama-3.2-1B}}

\vspace{2pt}

% ================= Row 1 =================
\includegraphics[width=0.249\textwidth]{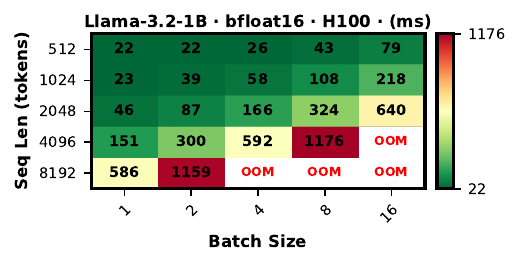}%
\hspace{-1mm}%
\includegraphics[width=0.249\textwidth]{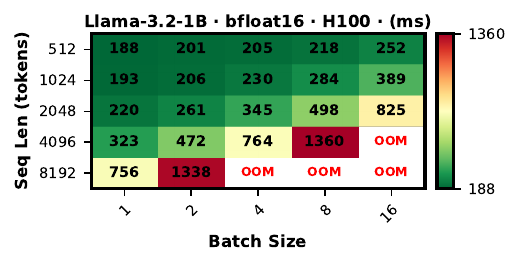}%
\hspace{-1mm}%
\includegraphics[width=0.249\textwidth]{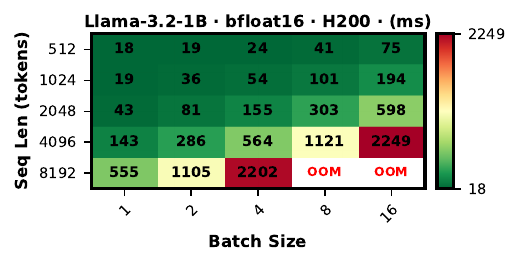}%
\hspace{-1mm}%
\includegraphics[width=0.249\textwidth]{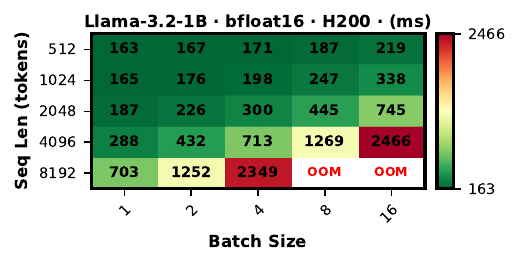}

\vspace{-3pt}

% ================= Model Label =================
{\footnotesize\textbf{Llama-3.2-3B}}

\vspace{2pt}

% ================= Row 2 =================
\includegraphics[width=0.249\textwidth]{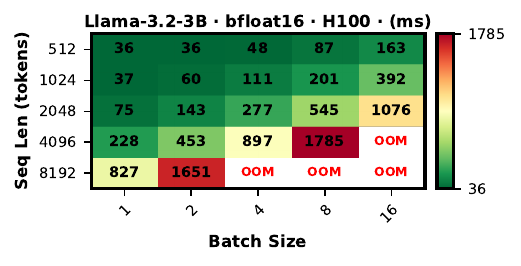}%
\hspace{-1mm}%
\includegraphics[width=0.249\textwidth]{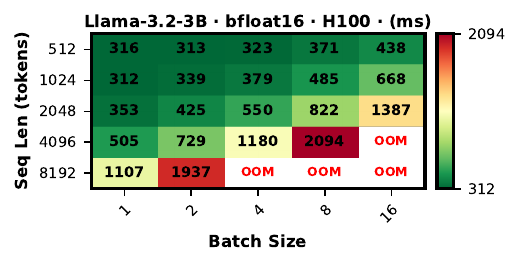}%
\hspace{-1mm}%
\includegraphics[width=0.249\textwidth]{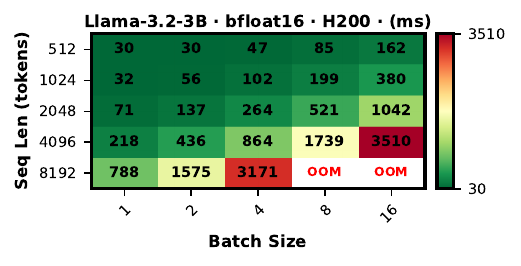}%
\hspace{-1mm}%
\includegraphics[width=0.249\textwidth]{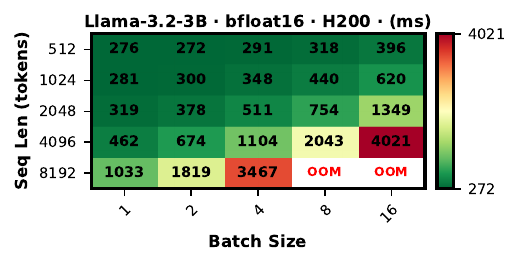}

\vspace{-3pt}
% ================= Model Label =================
{\footnotesize\textbf{OLMoE-1B/7B}}

\vspace{2pt}

% ================= Row 3 =================
\includegraphics[width=0.249\textwidth]{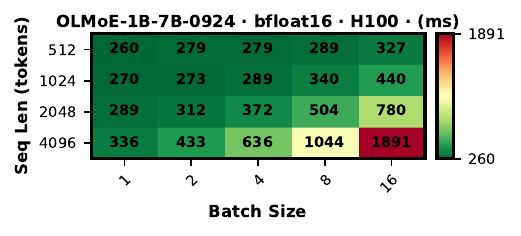}%
\hspace{-1mm}%
\includegraphics[width=0.249\textwidth]{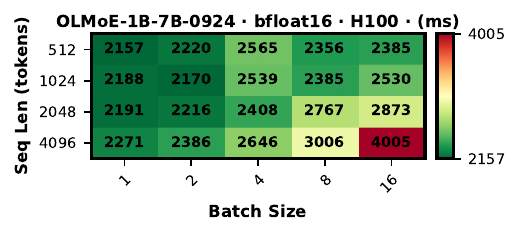}%
\hspace{-1mm}%
\includegraphics[width=0.249\textwidth]{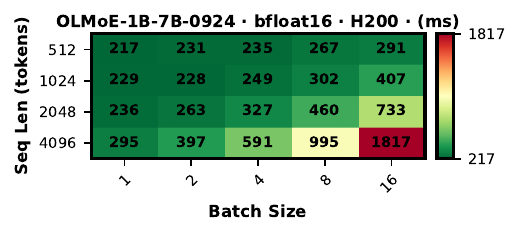}%
\hspace{-1mm}%
\includegraphics[width=0.249\textwidth]
{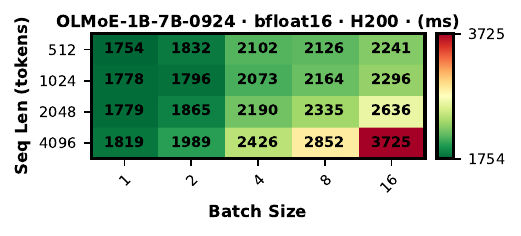}%
\vspace{-3pt}

% ================= Model Label =================
{\footnotesize\textbf{Qwen1.5-MoE-A2.7B}}

\vspace{2pt}

% ================= Row 4 =================
\includegraphics[width=0.249\textwidth]{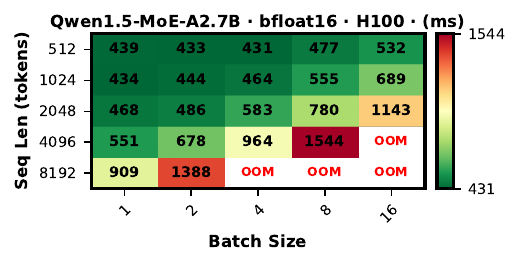}%
\hspace{-1mm}%
\includegraphics[width=0.249\textwidth]{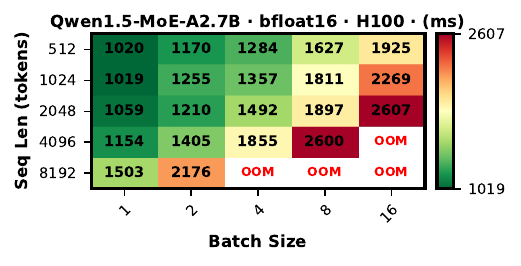}%
\hspace{-1mm}%
\includegraphics[width=0.249\textwidth]{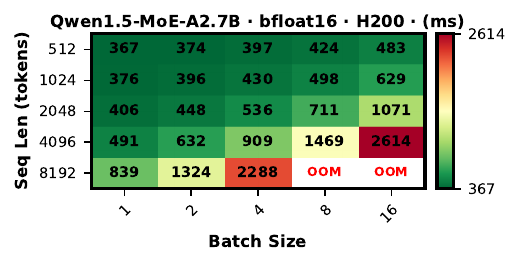}%
\hspace{-1mm}%
\includegraphics[width=0.249\textwidth]
{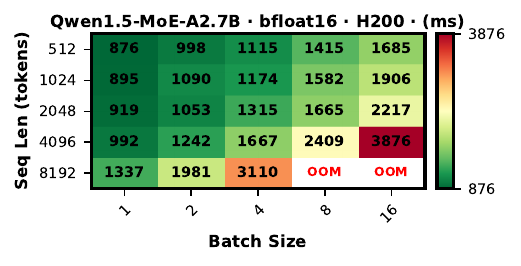}%

\caption{\textbf{End-to-end latency across dense and MoE LLM workloads.}
Heatmaps show end-to-end latency across batch sizes and input sequence lengths during prefill ($m=1$) and decode ($m=10$) on H100/H200 systems, where $m$ denotes the number of generated tokens. The decode heatmaps report total latency aggregated over a 10-token decode window. OLMoE-1B/7B does not support SL=8192 context length.}

\label{fig:e2e_sweep}

\end{figure*}

\subsection{End-to-End Prefill and Decode Latency}

We evaluate end-to-end inference in two phases: prefill (TTFT-oriented) and decode (aggregated latency over $m$ output tokens). Fig.~\ref{fig:e2e_sweep} reports latency across batch size (BS) and sequence length (SL) for dense and MoE workloads on H100/H200. Throughout all subsections, decode values correspond to the total over $m$=10 output-token decode run.

\textbf{Dense-Model Trends.} Dense prefill scales primarily with sequence length because attention work grows with prompt length: for Llama-3.2-1B on H100, latency rises from 22~ms at SL=512 to 586~ms at SL=8192 (26.6$\times$), while Llama-3.2-3B reaches 827~ms at SL=8192. Batching amortizes fixed overhead effectively in this regime: at SL=512, increasing BS from 1 to 16 raises Llama-3.2-1B prefill latency by only 3.6$\times$ (22~ms to 79~ms). Dense decode behaves differently. At (BS=1/SL=512), generating 10 decode tokens requires 188~ms vs. 22~ms for prefill over the same context. Unlike prefill, decode proceeds iteratively, so latency accumulates across repeated kernel launches, GPU execution, and KV-cache accesses. As sequence length grows, the decode bottleneck shifts toward KV-cache memory traffic, reaching 756~ms at SL=8192. Batch scaling again helps, primarily because device work grows: at SL=512, increasing batch size from 1 to 16 raises decode latency by only 1.3$\times$ while aggregate decode throughput improves 11.9$\times$ (53~tok/s to 634~tok/s), measured over the $m=10$ decode window across the batch.

\textbf{MoE-Model Trends.} MoE prefills and decodes differ qualitatively because routing introduces many short, dependency-chained kernels. On H100 at BS=1/SL=512, OLMoE-1B/7B takes 260.5~ms in prefill, 11.7$\times$ slower than Llama-3.2-1B despite similar activated parameter count, and dispatches 13,741 kernels vs. 850 for the dense model. Qwen1.5-MoE-A2.7B is even more fragmented at the same point (22,558 kernel launches). In decode, host bottleneck becomes dominant: OLMoE-1B/7B reaches 2157~ms at BS=1/SL=512 (11.5$\times$ slower than dense Llama-3.2-1B) and remains nearly flat across context length, varying only from 2081~ms (SL=512) to 2164~ms (SL=4096). This weak context sensitivity reflects a dispatch-dominated regime: at BS=4/SL=2048 ($m$=10), OLMoE-1B/7B dispatches 9,305 kernels per token (Table~\ref{tab:kernel-fragmentation}); with a per-kernel floor of about 4.7~$\mu$s (Table~\ref{tab:tsys_floor}), launch-path cost alone contributes about 44~ms per token before framework dispatch is added. The main levers are therefore reducing kernel count or amortizing the submission path.

\textbf{Cross-Platform Comparison.} H200 improves dense prefill most at short context ($\approx$18\%) and narrows toward $\approx$5\% at long context due to compute saturation. For decode, H200 delivers 12-15\% lower dense latency across sequence lengths and 14-19\% lower MoE latency. However, these gains do not remove the dominant MoE bottleneck: absolute latency improves, but routing- and dispatch-related host overhead remains first-order.

\textbf{Key Takeaway \#1.} \textit{Dense models shift from host-bound at small batches to compute- or memory-bound as workload grows. MoE decode does not: its latency stays dominated by host-side routing and dispatch, so hardware upgrades reduce absolute latency but do not change the bottleneck.}

\begin{figure}[!t]
  \centering
  \includegraphics[width=0.52\textwidth, height=6.4cm]{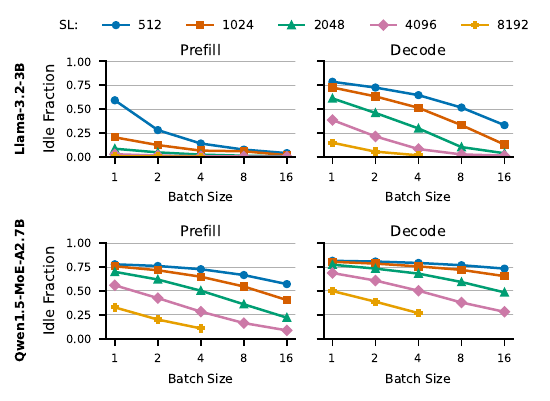}
  \caption{\textbf{Idle fraction across batch size and sequence length on H200}. The idle fraction indicates the portion of latency during which the GPU is not fully executing kernels.}
  \vspace{-16pt}
  \label{fig:idle_fraction_plot}
\end{figure}

\subsection{MoE Latency: Idle Fraction and Kernel Fragmentation}
\label{sec:idle_fraction}

While aggregate latency reveals that MoE decode is persistently host-bound, it cannot identify \emph{where} inference time is lost. We define the \emph{GPU idle fraction} (captures non-compute portion) as: $({(T_{\text{e2e}} - T_{\text{DeviceActive}})/T_{\text{e2e}}})$, where $T_{\text{e2e}}$ is the wall-clock inference latency. Fig.~\ref{fig:idle_fraction_plot} plots idle fraction across batch sizes and sequence lengths on H200, comparing between prefill and decode regimes.

\textbf{Llama-3.2-3B.} The dense model provides a baseline where orchestration overhead rapidly amortizes as workloads scale. During prefill execution, increasing sequence length or batch size expands the GEMM workload per kernel launch, causing the idle fraction to rapidly drop from 59.2\% (BS=1/SL=512) to 0.8\% (BS=1/SL=8192). Beyond $\text{BS}\geq4$ and $\text{SL}\geq2048$, idle fraction remains below 2.5\%, indicating that execution is effectively compute-bound. When decode is included ($m$=10), idle fraction rises due to repeated small-batch kernel bursts (78.5\% at BS=1/SL=512), but batching largely mitigates this effect, dropping below 5\% once $\text{BS}\geq8$ and $\text{SL}\geq2048$.

\textbf{Qwen1.5-MoE-A2.7B.} The MoE's idle fractions remain high across the sweep, indicating that inference does not become compute-bound even at large batch sizes and sequence lengths. In the prefill regime, the idle fraction at BS=1 ranges from 77.8\% (SL=512) to 32.8\% (SL=8192), comparable to the decode idle fraction of the dense model. Increasing the batch size only partially reduces this overhead: even at BS=16/SL=4096, Qwen1.5-MoE-A2.7B exhibits an 8.7\% idle fraction, compared to 0.2\% for Llama-3.2-3B under the same configuration (ratio of 44$\times$). This disparity persists during decode. At SL=512, increasing batch size from 1 to 16 reduces MoE idle fraction only from 81.5\% to 73.3\%. At BS=1/SL=1024, Llama-3.2-3B exhibits 20.5\% idle fraction while Qwen1.5-MoE-A2.7B reaches 75.7\% (3.7$\times$ higher).

\begin{table}[!t]
\centering
\small
\setlength{\tabcolsep}{6pt}   % adjust horizontal padding if needed
\renewcommand{\arraystretch}{0.90}
\begin{tabularx}{\columnwidth}{c *{4}{>{\centering\arraybackslash}X}}
\toprule
\textbf{Metric} &
\textbf{Llama-3.2-1B} &
\textbf{Llama-3.2-3B} &
\textbf{OLMoE-1B/7B} &
\textbf{Qwen1.5-MoE-A2.7B} \\
\midrule
Total kernel launches & 8,475  & 15,369  & 93,053  & 66,951 \\
Unique kernel names     & 77     & 76      & 222     & 223    \\
Kernels per token       & 847.5  & 1,536.9 & 9,305.3 & 6,695.1\\
Diversity ratio         & 0.0091 & 0.0049  & 0.0024  & 0.0033 \\
GPU utilization (\%)    & 58.9   & 67.6    & 15.5    & 27.7   \\
\bottomrule
\end{tabularx}

\vspace{4pt}
\caption{\textbf{Kernel fragmentation for dense vs. MoE models on H100 (BS=4/SL=2048, $m$=10).}}
\label{tab:kernel-fragmentation}
\vspace{-6pt}
\end{table}

Such patterns appear in single-GPU setting and cannot be attributed to inter-device expert communication. They point to non-compute overhead associated with expert routing and fragmented dispatch, which we next analyze using \name.

\subsection{TaxBreak Results}
\label{sec:taxbreak_validation}

\textbf{Kernel Fragmentation in MoE.} Table~\ref{tab:kernel-fragmentation} quantifies the structural source of MoE's persistently high idle fraction. At a fixed decode configuration (BS=4/SL=2048, $m$=10) on H100, OLMoE-1B/7B dispatches 93,053 GPU kernels per inference - 11$\times$ vs. Llama-3.2-1B (8,475) and 6$\times$ vs. Llama-3.2-3B (15,369). Qwen1.5-MoE-A2.7B similarly dispatches 66,951 kernels (8$\times$ Llama-3.2-1B). Normalized per output token, OLMoE-1B/7B requires 9,305 kernel launches per token vs. 848 for Llama-3.2-1B (an 11$\times$ difference). Despite this higher launch volume, the MoE \emph{kernel diversity ratio} ({unique names}/{total launches}) is \emph{lower}: 0.002 for OLMoE-1B/7B vs. 0.009 for Llama-3.2-1B, confirming that MoE fragmentation arises from many repeated invocations of a small set of routing and expert-GEMM kernels rather than from architectural heterogeneity. This launch explosion is consistent with OLMoE-1B/7B's 15.5\% GPU utilization (vs. 58.9\% for Llama-3.2-1B). Because dispatch is single-threaded, each additional kernel adds host-path cost; when kernels complete faster than dispatch, the GPU can become underfed. These results suggest that faster memory alone is unlikely to remove MoE’s host-bound behavior in such settings.

\textbf{Key Takeaway \#2.} \textit{At the measured decode configuration (BS=4/SL=2048, $m$=10), MoE models dispatch 8-11$\times$ more GPU kernels per output token than dense models of comparable active parameter count. This launch inflation stems from repeated invocations of a small set of routing and expert-GEMM kernels, which is not readily amortized by larger batches or faster memory. Unlike dense models, MoE inference remains persistently host-bound during decode, even at BS=16, since each token requires an order-of-magnitude more host-side dispatch work, which persists regardless of compute capacity or batch size.}

\textbf{$\mathbf{T_{sys}^{floor}}$ and $\mathbf{\Delta{KT_{fw}}}$ Characterization}. The null-kernel measurements in Table~\ref{tab:tsys_floor} indicate that $T_{sys}^{floor}$ is small and stable across Hopper platforms. We next ask whether real kernels launch close enough to that floor, for it to serve as a meaningful irreducible baseline. To answer that question, we decompose observed launch latency into the hardware floor $T_{sys}^{floor}$ and a residual framework-attributable term, $\Delta{KT_{fw}} = \max(0,\,T_{launch} - T_{sys}^{floor})$, and report the result by kernel family in Table~\ref{tab:kernel_launch_latency}.

\begin{table}[!h]
  \centering
  \setlength{\tabcolsep}{4pt}
  \small
  \begin{tabular}{lcccc}
    \toprule
    \textbf{GPU} & \textbf{avg} & \textbf{p50} & \textbf{p5} & \textbf{p95} \\
    \midrule
    H100 & 4.707 & 4.578 & 4.260 & 5.396 \\
    H200 & 4.503 & 4.452 & 4.177 & 4.909 \\
    \bottomrule
  \end{tabular}
  \caption{\textbf{Null-kernel $T_{sys}^{floor}$ ($\mu$s) measured in isolation.}}
  \label{tab:tsys_floor}
  \vspace{-10pt}
\end{table}

\begin{table}[!h]
\centering
\small
\setlength{\tabcolsep}{2pt}
\renewcommand{\arraystretch}{0.9}

\resizebox{\columnwidth}{!}{
\begin{tabular}{lcccccccc}
\toprule
& \multicolumn{4}{c}{\textbf{Llama-3.2-3B ($\mu$s)}} 
& \multicolumn{4}{c}{\textbf{OLMoE-1B/7B ($\mu$s)}} \\
\cmidrule(lr){2-5} \cmidrule(lr){6-9}

\textbf{Kernel Family} &
$p50$ & $p95$ & $\Delta KT_{fw}$ & $\uparrow$ &
$p50$ & $p95$ & $\Delta KT_{fw}$ & $\uparrow$ \\

\midrule

$\mathbf{T_{sys}^{floor}}$ (null)
& 4.75 & 4.75 & -- & --
& 4.73 & 4.73 & -- & -- \\

Scan (prefix)
& 5.07 & 5.16 & 0.32 & 7\%
& 5.12 & 5.13 & 0.40 & 8\% \\

Elem. (unroll)
& 5.11 & 5.28 & 0.36 & 8\%
& 5.11 & 5.20 & 0.38 & 8\% \\

Elem. (vector)
& 5.13 & 5.38 & 0.38 & 8\%
& 5.30 & 5.49 & 0.58 & 12\% \\

Reduce
& 5.30 & 5.30 & 0.55 & 12\%
& 5.07 & 5.46 & 0.34 & 7\% \\

Elem. (generic)
& 5.31 & 5.46 & 0.56 & 12\%
& 5.31 & 5.58 & 0.58 & 12\% \\

GEMM (nvjett)
& 5.93 & 18.58 & 1.18 & 25\%
& 5.57 & 6.17 & 0.84 & 18\% \\

GEMM (cuBLAS)
& 6.63 & 6.63 & 1.88 & 40\%
& 6.43 & 6.43 & 1.70 & 36\% \\

\bottomrule
\end{tabular}
}

\caption{\textbf{Per-family launch latency} (in $\mu$s) on H100 relative to the floor $T_{\text{sys}}^{floor}$ for Llama-3.2-3B and OLMoE-1B/7B (BS=1/SL=512; prefill). Most scan, reduction, and elementwise families remain close to the floor, while GEMM families show the highest $\Delta{KT}_{fw}$. The $T_{sys}^{floor}$ row reports the in-context replay floor, which differs slightly from the standalone measurement in Table~\ref{tab:tsys_floor} 
($\approx$0.04~$\mu$s) 
due to process-level CUDA context state.}
\label{tab:kernel_launch_latency}
\vspace{-8pt}
\end{table}

Table~\ref{tab:kernel_launch_latency} shows that most scan, reduction, and elementwise families launch close to $T_{sys}^{floor}$, with median residuals below 0.6~$\mu$s and typical deviations of only 7-12\% above the floor. This supports the use of $T_{sys}^{floor}$ as the irreducible launch baseline for the majority of kernel invocations. The main deviations occur in GEMM families: cuBLAS reaches 1.88~$\mu$s (40\% above $T_{sys}^{floor}$) for Llama-3.2-3B and 1.70~$\mu$s (36\%) for OLMoE-1B/7B, while nvJit GEMMs also sit above the floor. Thus, Table~\ref{tab:kernel_launch_latency} supports the intended split: $T_{sys}^{floor}$ captures the hardware floor common to all launches, and $\Delta{KT_{fw}}$ captures the family-dependent software excess above that floor. $p95$ value for Llama-3.2-3B \texttt{nvjet} GEMM family reaches 18.58~$\mu$s despite a much lower median ($p50$). We interpret this as a long-tail launch anomaly, likely reflecting variant-selection or framework/runtime replay effects rather than a change in the underlying floor.

% \begin{figure}[!h]
%   \centering
%   \includegraphics[width=0.49\textwidth]{new_fig/kernel_launch_latency_final_camera_ready.pdf}
%   \caption{\textbf{HDBI vs. TKLQT for GPT-2 (SL=512) across varying batch sizes on H100.} Regime characterization comparisons between HDBI and \cite{vellaisamy2025characterizing}.}
%     \vspace{-10pt}
%   \label{fig:HDBI}
% \end{figure}

\begin{figure}[t]
  \centering
  \captionsetup{skip=4pt}
  \begin{subfigure}[t]{\columnwidth}
    \centering
    \includegraphics[width=0.92\columnwidth,height=3.1cm]{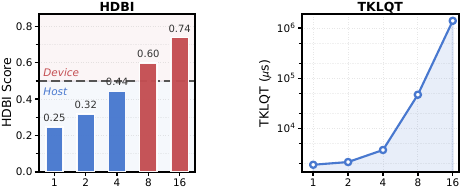}
    \caption{HDBI vs.\ TKLQT. HDBI crosses the host/device boundary near BS$=8$, while TKLQT rises sharply as the GPU becomes saturated in the same region.}
    \label{fig:HDBI_sub}
  \end{subfigure}

  \vspace{2pt}

  \begin{subfigure}[t]{\columnwidth}
    \centering
    \includegraphics[width=0.95\columnwidth,height=3.6cm]{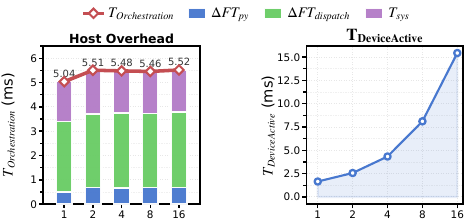}
    \caption{Host orchestration decomposition vs.\ device‑active time.}
    \label{fig:orchestrator_sub}
  \end{subfigure}

  \vspace{1pt}
  \caption{\textbf{Host vs device behavior for GPT‑2 (SL=512) across batch sizes on H200.}}
  \label{fig:combined_orchestrator_HDBI}
  \vspace{-10pt}
\end{figure}

\textbf{GPT-2 Case Study.} We use GPT-2 on H200 to compare \name against prior TKLQT-based characterization~\cite{vellaisamy2025characterizing}. Fig.~\ref{fig:HDBI_sub} shows that HDBI rises monotonically with batch size, from 0.25 (BS=1) to 0.74 (BS=16), placing the host-to-device crossover between BS=4 and BS=8.

\begin{figure*}[!t]
  \centering
  \includegraphics[width=\textwidth, height=5.4cm]{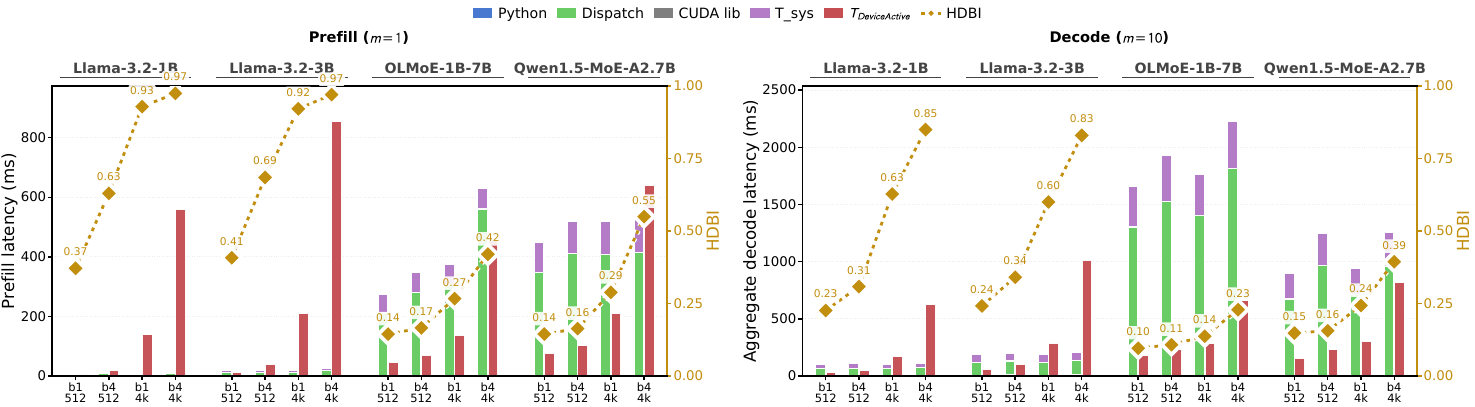}
  \caption{\textbf{H200 $T_{Orchestration}$ and HDBI characterization across dense and MoE workloads.} Stacked bars decompose $T_{Orchestration}$ into Python, dispatch, CUDA-library, and $T_{sys}$ contributions; adjacent bars show $T_{DeviceActive}$; diamonds show HDBI. Prefill uses $m$=1. Decode bars and HDBI points are totals aggregated over the matched $m$=10 decode run. \emph{Note: y-axis scales are different between panels; for cross-panel absolute-time comparisons use the HDBI diamonds with the same scale.}}
  \vspace{-8pt}
  \label{fig:hdbi_h200_broader}
\end{figure*}

Fig.~\ref{fig:orchestrator_sub} explains the source of that transition. $T_{Orchestration}$ stays nearly flat across batch sizes ranging from 5.04~ms (BS=1) to 5.52~ms (BS=16). This batch-size invariance is consistent with a serial dispatch path: larger batches increase GPU work without materially changing host submission behavior. Normalizing by kernel count (376 to 394 across batch sizes) yields a per-kernel host cost of ${\approx}$13.7~$\mu$s that is effectively constant (${\pm}$2.3\%), confirming that $T_{Orchestration}$ scales linearly with $N$ at a fixed per-dispatch overhead. Instead, the dominant change is in device work: $T_{DeviceActive}$ rises from 1.66~ms (BS=1) to 15.43~ms (BS=16). Within host decomposition, the H200 GPT-2 run is also notably stable: $T_{Py}$ increases only from 0.50~ms to 0.68~ms, $T_{dispatch\_base}$ stays within 2.89-3.11~ms, $\Delta{CT}$ remains zero for this workload, and the $T_{sys}^{floor}$ contribution remains nearly constant at 1.65-1.80~ms. This zero-$\Delta{CT}$ result reflects that GPT-2's matrix-multiply kernels are emitted as framework-native \texttt{nvjet}/\texttt{gemv2T} rather than cuBLAS/cuDNN, so $\mathbb{I}_{lib}=0$ and $\Delta{CT}$ is gated to zero.

This behavior also clarifies the relationship between HDBI and TKLQT (Fig.~\ref{fig:HDBI_sub}). Both metrics capture the same regime shift, but TKLQT rises much more sharply at larger batch sizes because it includes queue delay once the GPU becomes more occupied. HDBI remains interpretable across both regimes because it reflects the balance between $T_{DeviceActive}$ and $T_{Orchestration}$ rather than the exposed launch path alone. For GPT-2 on H200, \name shows that the crossover is driven by increasing device work while host orchestration remains nearly flat and is dominated by $T_{dispatch\_base}$, with no hidden vendor-library contribution.

\textbf{$\mathbf{T_{Orchestration}}$ and HDBI Characterization}. Fig.~\ref{fig:hdbi_h200_broader} illustrates the H200 characterization $T_{Orchestration}$, $T_{DeviceActive}$, and HDBI across sequence lengths and batch sizes. Decode values are aggregated over the $m$=10 autoregressive steps, hence the values reflect cumulative per-token execution.

The trend is a progressive divergence between dense and MoE execution as the workload moves from prefill into decode. At BS=1/SL=512 on H200, total $T_{Orchestration}$ for the $m$=10 decode window is 102.13~ms for Llama-3.2-1B and 188.27~ms for Llama-3.2-3B, compared with 10.5~ms and 17.6~ms, respectively, in single-step prefill ($m$=1). The per-step orchestration cost is nearly identical, with the decode totals implying ${\approx}$10.2~ms and ${\approx}$18.8~ms per step, respectively. This invariance is consistent with the kernel dispatch data: Llama-3.2-1B issues 850 kernel launches in prefill and 8,437 over ten decode steps (${\approx}$844 per step), showing that for a fixed dense architecture in eager mode, the dispatch count $N$ per forward pass is approximately shape-invariant. Therefore, the ${\approx}$10$\times$ growth in total $T_{Orchestration}$ reflects the multiplicative effect of $m$=10 steps. HDBI nevertheless drops sharply - from 0.37 to 0.23 for Llama-3.2-1B, and from 0.41 to 0.24 for Llama-3.2-3B - because per-step $T_{DeviceActive}$ is much smaller in decode (processing one new token per step) than in prefill (processing 512 tokens in parallel), shifting the balance toward host-visible execution.

That host-bound shift does not persist indefinitely for dense models. At BS=1/SL=512, the smallest decode case, dense models are most host-visible (HDBI$\approx$0.23). The larger decode cases at BS=4/SL=512, BS=1/SL=4096, and especially BS=4/SL=4096, move back toward device-dominant execution as per-step GPU work grows with batch size and context length. $T_{Orchestration}$ grows much more slowly than $T_{DeviceActive}$ across these points, and HDBI rises correspondingly. The dense-model trajectory is thus moderately balanced in prefill (HDBI$\approx$0.37-0.41), host-visible at small decode (HDBI$\approx$0.23-0.24), then increasingly device-dominant as batch and context scale.

MoE follows a different trajectory. It starts host-bound in prefill and remains host-bound as decode scales. At BS=1/SL=512 on H200, Qwen1.5-MoE-A2.7B and OLMoE-1B/7B prefill sit at HDBI$\approx$0.15, with $T_{Orchestration}$ far larger than $T_{DeviceActive}$ (448.8~ms vs. 75.9~ms for Qwen1.5-MoE-A2.7B; 273~ms vs. 46.1~ms for OLMoE-1B/7B). Decode at the same point ($m$=10) amplifies that gap sharply: Qwen1.5-MoE-A2.7B reaches 895.5~ms of $T_{Orchestration}$ with HDBI=0.15, while OLMoE-1B/7B reaches 1655~ms with HDBI=0.1. More importantly, unlike dense decode, the MoE decode cases at BS=4/SL=512, BS=1/SL=4096, and BS=4/SL=4096 still do not return to a balanced regime: host orchestration remains dominant and HDBI stays well below the dense-model values throughout the sweep. The dominant scaling trend is therefore not simply that decode is slower, but that MoE sustains host-boundedness over a wider region.

\textbf{Key Takeaway \#3.} \textit{$T_{Orchestration}$ and HDBI together reveal dense models become host-visible at small decode but return toward device-dominant as workload size grows, whereas MoE models remain host-bound from prefill to decode and persist at larger decode configurations. The key optimization depends on the trend in boundedness: reduce device work for dense long-context decode, but reduce routing- and launch-fragmentation overhead for MoE.}

\begin{figure}[!t]
  \centering
  \includegraphics[width=0.49\textwidth, height=3.7cm]{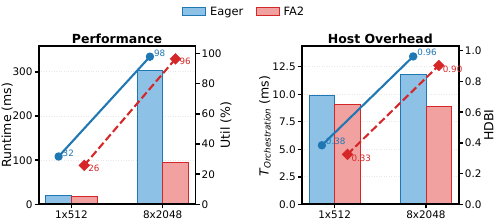}
  \caption{\textbf{Eager vs. FlashAttention-2 for Llama-3.2-1B on H200.} Bars show end-to-end runtime and $T_{Orchestration}$; lines show GPU utilization and HDBI for BS=1/SL=512 and BS=8/SL=2048.}
  \vspace{-12pt}
  \label{fig:fa2_h200}
\end{figure}

\textbf{Eager vs. FlashAttention-2.} Fig.~\ref{fig:fa2_h200} contrasts eager execution vs. FlashAttention-2~(FA2)~\cite{dao2023flashattention} for Llama-3.2-1B on H200 using \name. At smaller configuration (BS=1/SL=512), FA2 reduces end-to-end latency from 19.5~ms to 18.1~ms (7.2\%) and lowers $T_{Orchestration}$ from 9.8~ms to 9.1~ms (7.1\%). Despite improving absolute latency, FA2 removes device work faster than host overhead, which causes HDBI to decrease from 0.38 to 0.33 even as both $T_{Orchestration}$ and $T_{DeviceActive}$ improve in absolute terms.

At the larger configuration (BS=8/SL=2048), the performance impact is much larger: FA2 reduces end-to-end latency from 303.9~ms to 95.5~ms (68.6\%) while also reducing $T_{Orchestration}$ from 11.7~ms to 8.9~ms (24\%). GPU utilization remains effectively saturated in both modes (97.9\% eager vs. 96.5\% FA2), showing that the primary benefit at long context is not improved occupancy but a large reduction in device-active work per token - FA2's tiled, online-softmax kernel avoids materializing the full $N{\times}N$ attention matrix to HBM, substantially reducing memory traffic by eliminating intermediate read-back round trips. The HDBI decrease (0.96 to 0.90) reflects FA2 cutting $T_{DeviceActive}$ faster than host orchestration, enlarging the host-visible share despite both absolute values improving. This paired result illustrates the intended use of \name: a boundedness ratio alone can look counterintuitive after an optimization, whereas the decomposition shows that FA2 is succeeding mainly as a device-side optimization rather than a host-overhead reduction.

FA2 also exemplifies the kernel-fusion prescription from the \name diagnostic: by collapsing the multi-kernel attention sequence into a single fused kernel, FA2 reduces dispatch count from 850 to 791 ($-$7\%) at (BS=1/SL=512) and from 903 to 731 ($-$19\%) at (BS=8/SL=2048). The corresponding $N \cdot T_{sys}^{floor}$ term drops by 0.3~ms and 0.8~ms respectively, consistent with the eliminated launches $\times\;T_{sys}^{floor}$.

\textbf{Key Takeaway \#4.} \textit{FlashAttention-2 improves both host-visible orchestration and end-to-end latency, but the dominant win at larger sequence lengths comes from reducing device-side attention work rather than from eliminating host overhead. \emph{\name} and HDBI make this distinction explicit: FA2 lowers $T_{Orchestration}$ modestly, while a sharper reduction in GPU work drives the much larger runtime collapse.}

\section{CPU Single-Thread Performance Impact}
\label{sec:cpu_single_thread}

Eager-mode dispatch is single-threaded, so $T_{Orchestration}$ scales with kernel count $N$ on one CPU core. Our two platforms provide a useful controlled comparison. Both use Hopper-generation GPUs, but pair them with different host CPUs: H100 with an Intel Xeon 8480C (Sapphire Rapids, 2.0/3.8~GHz turbo) and H200 with an Intel Xeon Gold 6538Y+ (Emerald Rapids, 2.2/4.0~GHz turbo). H200 GPU runs at a lower  clock (1785~MHz vs.\ 1980~MHz, $-$9.9$\%$); end-to-end speedup cannot be attributed to faster GPU compute. The $T_{Orchestration}$/$T_{DeviceActive}$ decomposition allows us to separate host-dispatch effects from device-side effects with the same allocations (6 CPU cores, 32~GB, single GPU).

\begin{figure}[t]
  \centering
  {\scriptsize
  \textit{Llama = Llama-3.2-1B;
  Qwen = Qwen1.5-MoE-A2.7B;
  P = Prefill;
  D = Decode}
  \par}
  \vspace{1pt}
  \includegraphics[width=0.98\columnwidth]{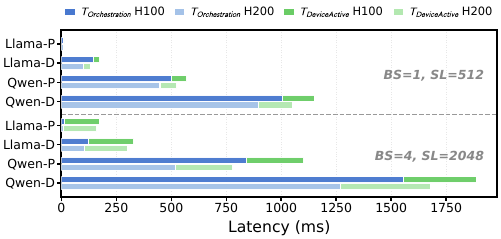}
  \vspace{-8pt}
  \caption{\textbf{Latency decomposition: H100 vs.\ H200.}
  Stacked bars show $T_{Orchestration}$ (blue) and $T_{DeviceActive}$ (green) for each H100/H200 pair. 
  Darker = H100; lighter = H200. 
  $T_{Orchestration}$ is consistently shorter on H200, while $T_{DeviceActive}$ is comparable or slightly longer. Decode is aggregated latency over 10 output tokens.}
  \label{fig:cpu_bars}
  \vspace{-10pt}
\end{figure}
Fig.~\ref{fig:cpu_bars} compares $T_{Orchestration}$ and $T_{DeviceActive}$  (BS=1/SL=512 and BS=4/SL=2048) for Llama-3.2-1B and Qwen1.5-MoE-A2.7B. Three findings:

\begin{enumerate}
    \item \textbf{Orchestration overhead tracks CPU single-thread speed.} $T_{Orchestration}$ is 10-29\% lower on H200. Since GPU is the same and H200 GPU clock is lower, this reduction is due to faster host-side dispatch on Emerald Rapids. $T_{DeviceActive}$ is comparable 
    %or slightly \emph{higher} 
    on H200, ruling out GPU memory bandwidth as the source of improvement. 

    \item \textbf{Decode amplifies CPU effect.} For Llama-3.2-1B, decode reduces $T_{Orchestration}$ by 29\% on H200, and 14\% in prefill, because decode issues  10$\times$ more kernel launches (8,437 vs.\ 850). Per-dispatch savings thus compound much more strongly in decode.

    \item \textbf{For host-bound MoE, CPU gains outweigh GPU penalties.} Qwen1.5-MoE-A2.7B on H200 is 8\% slower in ($T_{DeviceActive}$) yet 13-14\% faster end-to-end. With HDBI$\approx$0.12, $T_{Orchestration}$ is $\approx$7$\times$ larger than $T_{DeviceActive}$, so host-side improvement dominates the device-side loss. Dense Llama-3.2-1B sees smaller end-to-end gains because it is less host-bound.
\end{enumerate}

\begin{figure}[t]
  \centering
  {\scriptsize
  \textit{L = Llama-3.2-1B;
  Q = Qwen1.5-MoE-A2.7B;
  P = Prefill;
  D = Decode; \\
  Trailing digits = BS/SL}
  \par}
  \vspace{1pt}
  \includegraphics[width=\columnwidth]{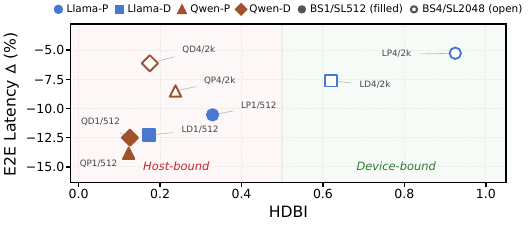}
  \vspace{-18pt}
  \caption{\textbf{E2E latency gain (H100$\to$H200) vs.\ HDBI.}
  Marker shape encodes model$\times$phase; filled is BS=1/SL=512, hollow is BS=4/SL=2048. Host-bound points show largest gains; device-bound points show smaller gains. Decode is aggregate latency over 10 output tokens.}
  \label{fig:cpu_scatter}
  \vspace{-8pt}
\end{figure}

This relationship holds at higher load, but its effect is gated by HDBI (Fig.~\ref{fig:cpu_scatter}). Llama-3.2-1B prefill at BS=4/SL=2048 is strongly device-bound (HDBI=0.93), so a 13\% drop in $T_{Orchestration}$ yields only a 5\% end-to-end gain. Qwen1.5-MoE-A2.7B at the same point remains host-bound (HDBI=0.24), and reducing $T_{Orchestration}$ still gives an 8\% latency gain. The pattern follows in decode: host-bound points benefit most from the faster CPU, while device-bound points see attenuated gains because $T_{DeviceActive}$ dominates.

\textbf{Key Takeaway \#5.} \textit{For host-bound LLM inference, especially MoE decode, CPU single-thread performance is a first-order design parameter. A faster host CPU reduces $T_{Orchestration}$ by 10-29\% and can improve end-to-end latency even when paired with a slower-clocked GPU. The effect weakens as the workload becomes more device-bound.}

\section{Conclusion and Outlook}

This work presents the \name methodology, a trace-driven decomposition of host-visible inference overhead into framework translation, CUDA-library translation, and launch-path components. Across dense and MoE LLM workloads on NVIDIA H100 and H200 systems, the results show that previous approaches are insufficient to identify key performance impediments and potential optimization targets. \name can identify whether the key opportunity lies in the software stack, the launch path, or the device-side. The HDBI provides a diagnostic boundedness metric to help guide optimization.

This study has some limitations. The methodology is evaluated on NVIDIA GPUs and is dependent on CUDA tracing interfaces. 
Also, replay-based attribution can be imperfect for highly dynamic, autotuned, or synchronization-heavy kernels. 
%despite the pipeline's matching and fallback safeguards. 
In addition, HDBI is a diagnostic work ratio, not an optimization objective in itself, and should be interpreted alongside absolute latency and \name decomposition.

For host-bound workloads, CPU single-thread performance is a first-order parameter: faster single-core execution reduces $T_{Orchestration}$ by 10-29\%, whereas additional cores provide no benefit. The cross-platform (H100 vs. H200) comparison (Section~\ref{sec:cpu_single_thread}) shows that for MoE inference, a faster host CPU can improve end-to-end latency by 11-14\% even when paired with a slower-clocked GPU. This effect persists across prefill and decode, and weakens as HDBI rises above $\approx$0.3.
Our future work will focus on more diverse AI workloads and platforms (NVIDIA GB200/GB300 and AMD MI300A/MI300X).

\bibliographystyle{plain}
\bibliography{references}

\end{document}